\documentclass[aps,twocolumn,twoside,floatfix,nofootinbib,a4paper]{revtex4}
\usepackage{graphicx}
\usepackage{marginnote}

\usepackage{hyperref}
\usepackage{enumitem,kantlipsum}
\usepackage[usenames,dvipsnames]{color}
\usepackage{amsmath}
\usepackage{amssymb}
\usepackage{amsfonts}
\usepackage{mathrsfs}
\usepackage{bm}
\usepackage[all]{xy}
\usepackage{hyperref}
\usepackage{tikz-cd}

\newcommand{\beq}{\begin{equation}}
\newcommand{\eeq}{\end{equation}}
\newcommand{\bea}{\begin{eqnarray}}
\newcommand{\eea}{\end{eqnarray}}
	
\newcommand{\ea}{\end{align*}}

\newcommand{\bma}{\begin{pmatrix}}
\newcommand{\ema}{\end{pmatrix}}

\usepackage[percent]{overpic}
\usepackage{overpic}

\begin{document}
\title{Particulate exotica}
\author{Fan Zhang} 
\affiliation{Gravitational Wave and Cosmology Laboratory, Department of Astronomy, Beijing Normal University, Beijing 100875, China}
\affiliation{Advanced Institute of Natural Sciences, Beijing Normal University at Zhuhai 519087, China}

\date{\today}

\begin{abstract}
\begin{center}
\begin{minipage}[c]{0.9\textwidth}
Recent advances in differential topology single out four-dimensions as being special, allowing for vast varieties of exotic smoothness (differential) structures, distinguished by their handlebody decompositions, even as the coarser algebraic topology is fixed. Should the spacetime we reside in takes up one of the more exotic choices, and there is no obvious reason why it shouldn't, apparent pathologies would inevitably plague calculus-based physical theories assuming the standard vanilla structure, due to the non-existence of a diffeomorphism and the consequent lack of a suitable portal through which to transfer the complete information regarding the exotic physical dynamics into the vanilla theories. An obvious plausible consequence of this deficiency would be the uncertainty permeating our attempted description of the microscopic world. We tentatively argue here, that a re-inspection of the key ingredients of the phenomenological particle models, from the perspective of exotica, could possibly yield interesting insights. Our short and rudimentary discussion is qualitative and speculative, consisting merely of conjectures, because the necessary mathematical tools have only just began to be developed. 
 \end{minipage}
 \end{center}
  \end{abstract}
\maketitle

\raggedbottom
\section{Introduction}
A long lineage of geometrodynamical investigations has evolved over time (see e.g.,\cite{Geometrodynamics,1957AnPhy...2..525M,PhysRev.97.511,2004gr.qc.....9123A,Atiyah:2011fn}). Ever since the early days of modern particle physics \cite{Kelvin}, there had always been the temptation to connect the observed discreteness and conservations of particle physics with nonlocal features of various geometric patterns formed out of the fabric of spacetime itself, due in part to the aesthetics of the emulsion of particle physics with General Relativity. Also, this way, the particles would automatically be welded onto our universe and can never move off of it. There is no need then to either compactify any extra dimensions into tiny sizes to make them invisible at low energies, and suffer the associated landscape problem \cite{CANDELAS198546}, or to conjure up some sort of confining potential to assemble a thick brane \cite{RUBAKOV1983136}. 

In order to carry out the geometrodynamics program, one could enlist tools, in ascending order of finesse, that describe a 4-dimensional shape: 
\begin{enumerate}
\item Its algebraic topological (sometimes abbreviated to just ``topology'' below) characteristics. Equivalence between two manifolds at this level of detail is established by homeomorphisms. 

\item Its smoothness structure, the classification of which is established with diffeomorphisms. 

\item Its complex structure, classified by biholomorphisms. 

\item Its (pseudo-)Riemannian geometry, distinguished by isometries. 
\end{enumerate}
Historically, efforts have concentrated on the top and bottom lines, letting geometry handle continuum features such as energetics, and leaning on algebraic topological features (e.g., various characteristic classes) to generate discrete properties. 
In this note, we advocate evoking the more fine-grained second entry in the list to assist with the latter task, in order to procure additional flexibility for nuances, from exotic smoothness\footnote{In principle, we could also have different complex structures on top of a fixed smoothness structure, so evoking nontriviality in complex structures might also buy us something. However, since assuming the wrong complexity will mostly just break holomorphy and cause complex conjugations to appear, we would only really be missing vital information if our particle theory is presently constricted to consist solely of holomorphic expressions, which is not the case. In other words, correcting for nontriviality in complex structures will at most grant us more convenience and mathematical elegance, but not core modeling competence. This task is less urgent and so we will not execute it here. Also, we do not exclude incidental nontrivialities in algebraic topology, but they are no longer the main genesises of particle features.}. 

The ``fine-grain'' qualification comes from the fact that the smoothness consideration can be seen as a further tightening of topology that tames its wilder beasts and refines its rather broad-stroked equivalences. The definition of topological equivalence, that there exists homeomorphisms (continuous reshuffling of points), was meant to convey the directive that we can morph one shape into another, in a fashion similar to stretching and squeezing playdough, but without tearing or pinching it. However, more recent developments on objects like fractals\footnote{The fractal example is only a particular(ly popularized, thus accessible) case of a more subtle and rich issue.} alert us to the fact that the mathematical definition of continuity is perhaps rather more lenient than we initially intended\footnote{It is possible for a collection of singletons (cf., the Cantor set) to avoid being isolated -- i.e., to always have friends arbitrarily nearby -- making it suitable to serve as the codomain of a continuous function (e.g., a homeomorphism). Yet it still remains disconnected -- each singleton is always separated from any friend by some ``lava'' in-between -- thus we need to hop over the lava when tracking the image of that function, potentially exploding formal derivatives.} (cf., the Cantor function), allowing homeomorphisms to shuffle points in a haphazard fashion that one would not have expected based on the playdough intuition. Namely, it is permitted to rough an originally smooth manifold up into becoming rather pathological (e.g., contains regions that are nowhere differentiable). And even if we demand that the overall resulting manifold be smooth, the originally smooth submanifolds could still suffer the battering and become wildly embedded slices\footnote{As an aside, we mention that the wild embedding into spacetime can be seen as a generalized knotting since wildness is reflected in the homotopy and/or homology of the embedding complement \cite{Embedding}.}.  
To close this loophole and get back to more familiar grounds, one could specialize to the more restrictive smooth category (or the piecewise-linear category if some mild kinks can be tolerated, which is in any case equivalent to the smooth category for four dimensions), wherein equivalence is established by the gentler diffeomorphisms, or everywhere (usually infinitely, but see Theorem 1.1.6 in \cite{KirbyCal} and reference therein) differentiable one to one mappings. 

Many authors (see e.g., references in \cite{ExoticSmooth}) had perceptively already began exploring this exciting new territory, and we follow in their footsteps. Since four dimensions is uniquely susceptible to hosting vast multitudes of exotic smoothness structures\footnote{See e.g., \cite{FourMani} for a summary of why 4-D is exceptional. Explicitly, there are worked out examples of exotic $\mathbb{R}^4$ \citep{gompf1985} that is unique for 4-D \cite{stallings_1962}, as well as exotic $S^3\times \mathbb{R}^1$, $D^2\times \mathbb{R}^2$, $S^2\times S^2$ \citep{2010arXiv1005.3346A} and the Mazur manifold \citep{10.2307/1970288,BSMF_1960__88__113_0}, and we also have the possibility for exotic $S^4$ (cf., the still open smooth 4-D Poincar\'e conjecture, see in particular \cite{Yasui2015NonexistenceOS}). However, we lack algorithms for systematically enumerating exotic structures of any arbitrary topology, or laying down actual exotic coordinate charts. Such gaps in knowledge seriously hamstring attempted quantitative investigations on the implications of exoticity for physics.}, this type of proposals thus effectively explain why we live in the spacetime dimension that we do, via an anthropic argument.   
The technical details of our proposal differ from previous literature though, and follow instead the codimension-one\footnote{Equiaffine metric being a second fundamental form takes value in $\mathbb{R}^1$, implying that the conormals form a line bundle and there is only one real codimension. Also, there isn't a Ricci-K\"uhne equation handling the normal fundamental form.} \footnote{Some properties like non-compactness or non-orientability tend to require higher embedding dimensions, but exoticity doesn't appear to. 
E.g., some exotic $\mathbb{R}^4$s can be embedded into standard $\mathbb{R}^4$ while the rest (``large'') can be embedded into $\mathbb{R}^5$. See further e.g., \cite{2020arXiv200603109K} for the insensitivity (and caveats) of embedding calculus to exotic differential structures.} braneworld scenario of \cite{galaxies8040073}. 
In particular, we take the view that the specific shape of our spacetime is a solution of some isoperimetric problem\footnote{Cf., \cite{galaxies8040073}, if one envisions our spacetime to be a membrane residing within a higher dimensional ambient, perhaps as the domain wall separating two different bulk phases, then the internal tension within the membrane would prefer a minimized area. In this sense, we live in a world akin to that studied by surface condensed matter theory, but with an even more interesting surface dimension.}, which prefers smoothness since kinkiness can increase (in the extreme case of wildness, explode to infinity), e.g., the surface area of a bubble, without changing the enclosed volume. This braneworld scenario thus underpins the implicit assumption that our spacetime carries a smoothness structure in the first place, and also piques our interest in closed manifolds, in contrast to the previous focus on open exotic $\mathbb{R}^4$s, whose exoticity arise for different reasons and behave quite distinctly. 
Other points of departure from previous literature include, e.g., we do not envisage the exotic core region corresponding to particle worldlines to satisfy vacuum Einstein equations like in \cite{Brans:1992mj,Brans:1994hq} (i.e., the cubic form of \cite{galaxies8040073} does not vanish). Also, one particle only corresponds to a single exotic smoothness structure, instead of a legion of them each being one quantum state\footnote{These are geared more towards quantum gravity, turning the many worlds interpretation quite literal -- each world has a different spacetime topology or smoothness structure. We will view quantum-ness as being less fundamental than this, and being due to limitations in our modeling sophistication instead.} as in \cite{AsselmeyerMaluga:2010ti,2020Symm...12..736P}. 

Beyond furnishing more versatility into our modeling toolbox, exotica also  renders plausible apologias for some of the shortcomings in our current effort. Because model builders had tried to describe the physical world using differential equations, the smoothness structure of the spacetime, and not just its topology, is particularly relevant for physics. 
While it could be considered (notwithstanding our isoperimetric argument) a matter of debate as to whether the assumption of smoothness is appropriate, or if spacetime is fundamentally granular, the differential equations strategy seems to have done a reasonable job so far, encountering only a few difficulties, e.g., when it comes to interpreting quantum mechanics. The mild symptoms then appear to suggest that the disease is not that spacetime is not differentiable at all, but possibly a less fatal one, that we are using a wrong smoothness structure near some spatially compact regions surrounding the microscopic particles. In the differential topology lingo, the spacetime is exotic, but not fake (we also use the term ``wild''), and the ills infesting and plaguing our theories arise out of our ignorance of the exoticity\footnote{That we'd assume a wrong differential structure is easy to understand. Our intuitions are based off of our daily experiences in three spatial dimensions, where there is no exoticity (cf., Moise's theorem \cite{10.2307/1969769}). In fact, the discovery of the very first exotic manifold (in 1956, well after the foundations of particle theories had already been established), was a surprise to mathematicians.}.  
Specifically, missing exoticity will cause the factually smooth (under the physical exotic structure) quantities to appear erratic and non-differentiable (under the vanilla structure we build models on), thus incapacitating the derivatives in our theories. To mitigate the problem, we consciously or unwittingly resort to imitating stochastic calculus, which is designed to manipulate similarly (but not similar enough, thus all the residual perplexity, we will elaborate in Sec.~\ref{sec:equations} below) non-differentiable entities. But we pay a heavy price, not least in predictability, the sacrifice of which renders our theories quite less effective, since arguably the whole purpose of physical theories is to be able to foretell future evolutions given initial data. 

To properly mourn our losses, we begin by recalling that one of the consequences of exoticity is that na\"ive topologically trivial spatial slices like $\mathbb{R}^3$ or $\mathbb{S}^3$, that we usually adopt in physics, must actually be fake\footnote{The fakeness of the spatial slices are stabilized by multiplying with time since the entire spacetime is still smooth, albeit exotic -- the spatial slices are thus inhibited in their wildness and must be manifold factors \cite{10.2307/1970322}.} \footnote{The na\"ive slices also miss out on some topological features relating to fermions, cf., the lair discussions in Sec.~\ref{sec:particlestop}. However, they can be seen as providing a legitimate topological (although not smooth) foliation of the macroscopic universe after we pinch out the lairs (see Sec.~\ref{sec:fields}) to arrive at a coarse-grained portrayal of the particles.}, or else their and the temporal direction's (unique and standard) smoothness structures would direct-multiply into the standard plain vanilla differential structure of the whole spacetime. Equivalently, one could note that the mapping between a vanilla and an exotic spacetime would be a homeomorphism that is not a diffeomorphism, meaning that, as alluded to in the last paragraph, some vanilla-ly smooth submanifolds (i.e., the spatial slices) would be ruffled into wild images. Yet another way to intuit the situation is by noting that manifolds sharing the same topology, but carrying different smoothness structures must actually be quite different shapes from a smooth embedding point of view (imagine fleshing out the exotically smooth Kirby diagrams in \cite{AkbulutBook} into embedded handlebodies), because it is this difference that subsequently become internalized into the distinction in smoothness', which the two manifold both inherit from the same ambient differential structure.
Consequently, we shouldn't expect to be able to smoothly foliate them in the same way. 

The spatial slices being fake would prevent them from serving as legitimate Cauchy surfaces, thus even if we had recognized the possibility and implications of exoticity, and managed to write down correct differential equations of motions for the particles (i.e., the detailed geometric equations for the spacetime, such as the structure equations for moving frames, see e.g., \cite{1951tgfc.book.....C,CartanBook}), we still wouldn't be able to solve the initial value problem and have a predictive physical theory. Worse still, we didn't anticipate the wildness of the spatial slices precisely because we didn't expect exoticity, and thus stand no chance of getting the differential topology of the core regions of the particles right, let alone the equations describing the more detailed geometry there. We have to at best rely on a coarse-grained description that evolves a redacted equation of motion, which must be further randomized to handle the wildness. The redaction is epistemological in nature, while the wildness is more ontological, yet both contribute to the degradation of predictability, so they become difficult to disentangle in quantum theories, which inevitably become philosophically quite convoluted and confusing. 

Having these observations in mind, we believe that the acknowledgment of exoticity offers an enticing prospect for explaining away the more unusual features of quantum and particle physics. In particular, this approach folds the microscopic world back into our comfort zone of super-deterministic\footnote{The negative feelings against super-determinacy relies largely on the implausibility of collusion between the experimental apparatus (sometimes constructed to take inputs from distant quasars) and test particles. We note though that in the absence of an actual collapse of a physical wave function (not present in the popular Everett interpretation; and itself suffering from issues such as the observer dependence of time simultaneity), there is no non-statistical intrinsic arrow of time (even temporal reflection asymmetry won't necessarily provide us with one, if evolution can proceed either way without having to change the equations), so intuitions gleaned from observing systems evolving backwards in time are just as valid, and from this vantage point the strong collusion seems inevitable given that the post-measurement histories of the apparatus and test particles must by definition be tangled up (see also discussions in \cite{2021arXiv210304335H}).} objectivity
(and thus evades Bell's inequality, since all the conditional probabilities involved in its derivation trivializes), more aligned with the macroscopic theories exemplified by General Relativity. Furthermore, despite the use of the word ``exotic'', we are really trying to banish the more unruly wild beasts of mathematics from physical reality, so we see our proposal as being a conservative one. 

\section{Dictionary} \label{sec:Dic}
The go-to tools for investigating exoticity are the handlebodies (see e.g., \cite{KirbyCal,AkbulutBook}), which to smoothness are similar to how CW-complexes are to homotopy\footnote{A handle decomposition determines a relative cell complex with the same homotopy type, thus can be seen as a fleshed out version of the latter that sees finer details like smoothness structures. On the other hand, a handle attaching map is always smoothable in 4-D, so handles cannot deal with fake manifolds -- they are quite dedicated to differential structures. }. 
Intuitively, handles are an attempt at breaking down the problem of studying a manifold into smaller puzzles by examining its standardized constituent pieces, and is thus similar to local charts in this respect, and the detailed arrangement governing how handles connect up into the whole manifold\footnote{That is, the homotopy and isotopy of the attaching spheres -- the latter measures knottedness thus only active for 2-handles and above that have attaching spheres with sufficient dimensions to be knotted, the knotting and linking of 2-handles' attaching spheres is a major source of exoticity for 4-manifolds.} is akin to how charts join up to form an atlas (inequivalent ways result in different smoothness structures), but is more accessible since it does not need detailed coordinates to be laid down, which are vital if we are dealing with differential geometry but mere spurious appendages if only differential topology is being probed. 

If we wish to identify particles with concentrations of exoticity, then an obvious correspondence presents itself: a composite particle or a collection of particles participating in a scattering experiment can be seen as a complicated handlebody, and our attempt at describing them in terms of more rudimentary elementary particles is in essence a handlebody (partial) decomposition exercise. In particular, our entire universe with its many particles is just one extraordinarily complicated handlebody, and the fact there are typically infinitely many (countably so for closed manifolds, and uncountably so for open ones) different exotic differential structures for four dimensional manifolds is conducive to accommodating the large particle population. 

However, one must be cognizant of the fact that handles are four dimensional, while particle decomposition is only carried out over three dimensional spatial slices, so two particles well-separated at some time but comes into interaction later on won't represent handlebody subunits that are disjoint. In particular, just like an ever-lasting stable particle is a reductive conceptual tool that likely won't exist in the actual universe, its corresponding handlebody is likely not a constituent of the actual universe. As well, it is important to be aware that handlebody decomposition is not unique, although once detailed geometric considerations are also included, providing us with a spatial metric, distance-based discriminant setting apart alternative groupings of handles should become attainable. 

\subsection{Particles} \label{sec:particles}
\subsubsection{Topological identity} \label{sec:particlestop}
We take cues from spin statistics. The indispensable employment of the double cover $\text{SL}(2,\mathbb{C})$ of the Lorentz group $\text{SO}^+(1,3)$, and spinors, forming a projective representation of the latter (although it is called a representation, it is an illegitimate one, since given an element from $\text{SO}^+(1,3)$, the reaction of spinors is not unique), signifies that, within our geometrodynamic context, there are likely complicated hidden crevices of spacetime, the responses of which to the isometries in the visible macroscopic sector outside are not currently being properly understood. Intuitively, one may call up the Balinese cup trick (or the Dirac's string trick), for which the central point is that, under the candle dance routine (no feet manoeuvre), the dancer, also being a part of the overall ``spinning system'', doesn't return to the original orientation if just the cup spins one cycle. The ``hidden crevice'' we seek is this dancer that the audiences (macroscopic observers), fixated on the cup, perfunctorily notice (via the physical consequences of half integer spins), but fail to fully account for. The role of the differential structure in this analogy is then to provide the necessary rigidity (as compared to just topology) to restrict how the dancer's joints are allowed to twist. 

In other words, the spacetime geometry needs to be augmented with further nontriviality, e.g., by adding extra features via connect-summing. Note, since such features are clearly not being mapped out in detail by our present theories, meaning that coordinates and metrics inside them, unlike those outside, never explicitly appear in the quantum field expressions (i.e., they are ``hidden'' crevices; we peep inside by sampling a bare minimum of interior information before grafting it onto points in the connecting region exposed to the outside world, see discussions below), thus there is no recipients to subject isometries onto. In other words, we do not need $\text{SL}(2,\mathbb{C})$ to be an actual isometry inside said features in order to retain compatibility with existing theories. It simply being a continuation of the $\text{SO}^+(1,3)$ transformations in the macroscopic outside world would suffice. 

Given that the fermions serve as the fundamental building blocks of the physical material world, their dictionary entry in the book of exotica should similarly be the ``atoms'' of topology for 4-manifolds. We propose that at the root level (additional further structures will be attached to further fashion the moderately diverse fermion botany), these are the complex projective plane $\mathbb{C}\mathbb{P}^2$ and its orientation reversed mirror image\footnote{It is more commonly denoted $\overline{\mathbb{C}\mathbb{P}}^2$ in literature, but that notation may lead to confusions with complex conjugation, which preserves the orientation since there are two complex dimensions.} $-\mathbb{C}\mathbb{P}^2$, which are quite appropriately rudimentary ingredients\footnote{\label{fn:legos}These are somewhat similar in functionality to the $\mathbb{RP}^2$s in the classification of 2-manifolds. In other words, they are base units of nontrivial topology, and their total population (i.e., the overall number of fermions and antifermions) is the rank of the intersection form, which is a topologically important index.} in the sense that one can build any orientable closed\footnote{Physically, if the universe has a boundary, explaining the boundary conditions there would present new challenges, so we do not assume so. This is also mathematically convenient as a handlebody for a non-compact manifold can become much more complicated. Moreover, we demand orientability for our universe, in order to have a spin structure.} smooth (including the exotic ones) 4-manifold by applying a four dimensional generalization of the Dehn surgery along a $T^2$-link on a raw initial dough\footnote{``Surgery'' largely refers to disfigurements that involve the pinching, cutting or suturing type of operations that are forbidden in homeomorphisms or diffeomorphisms (whatever equivalence category is being studied), and are thus utilized to create new and different (inequivalent) entities. We only concentrate on those that change the smoothness structure, so the algebraic topology and its ``atom'' population size should be largely preserved.}, which in turn is a connected sum of several copies of $S^1\times S^3$s and $\pm \mathbb{C}\mathbb{P}^2$s (see the appendix in \cite{Iwase}). Furthermore, these topological bricks also double up as elementary handlebodies. Specifically, attaching $-\mathbb{C}\mathbb{P}^2$s represents the blowing up of the spacetime\footnote{It is worth noting that the complex orientation of the ambient $S^1\times S^3$ (see Sec.~3.4 of \cite{KirbyCal}) is consistent with $-\mathbb{C}\mathbb{P}^2$ since connected sum involves an orientation reversal, so blow-ups normally use $-\mathbb{C}\mathbb{P}^2$ and not ${\mathbb{C}\mathbb{P}}^2$. Indeed, blow-up is a procedure originating from complex analysis, which thus provides an opening for complex numbers to infiltrate particle models. However, these models are not complex analytic, so while we utilize complex structures, we don't really need to perfectly comply with it, thus blowing-ups with $\mathbb{CP}^2$ are allowed, and anti-particles can exist. The discussions in this note applies to the anti-particles in much the same way as they do the particles, even when we neglect to explicitly state so in the interest of succinctness.}, which is equivalent to attaching a 2-handle with $-1$ framing\footnote{These framing numbers ($-1$ for a $\underline{\mathbb{CP}}^2$ and $+1$ for ${\mathbb{CP}}^2$) are also the second Betti number being added to spacetime due to the presence of a lair.} (see example 4.4.2 of \cite{KirbyCal}). Therefore, in what follows, we shall term the $\pm \mathbb{C}\mathbb{P}^2$ regions ``lairs'', and propose that they host anti-fermions and fermions. Note, we used the term ``host'' to emphasize that after the $T^2$-surgery that brings in exoticity, the web of lairs becomes intricately decorated with additional 2- and 1-handles\footnote{As per standard convention, we will not explicitly track the 3- and 4-handles, since they just fill in the gaps between the 1- and 2-handles, and are thus uniquely determined by them.} \footnote{In the Kirby diagram, the 1-handles are represented as a pair of spheres, where the attaching circles of 2-handles enter into one and exit through the other as if it had traversed a wormhole (or in Akbulut's dotted circle notation, the two spheres can be seen as collapsing and merging into a single disk that is bounded by the dotted circle, so the 2-handle attaching circles threading through the center of the dotted circle, hitting the disk, are those that tunnel through the 1-handle, entering into one mouth -- the disk -- and then exit immediately out of the other -- also the disk; this contraction of the tunneling section in the diagram makes it easier to see the knotting and linking of the 2-handle attaching circles). Geometrically, the tunneling is because a 2-handle can ride onto the 1-handle, so some segment of its attaching circle sits on the boundary of the 1-handle, not the boundary of the base 0/4-handle that the Kirby diagram depicts. Mathematically, the 1-handles are quite powerful at changing the characteristics of differential structures, e.g., the minimum number of 1-handles (among all possible alternative handlebody decompositions of the same manifold) is an invariant of differential structures \cite{2008arXiv0806.3010A}. Physically then, we should expect that adding 1-handles would engender new curiosity, e.g., the 1-handle wormholes may plausibly connect up the interiors of different fermions, allowing their guts to become intertwined, thus bind them together and possibly help explain phenomenons like quark confinement.} that either attach to the boundaries of individual lairs\footnote{In the Kirby diagram, these would be slid off onto the $S^3$ boundary of the base 0- or 4-handle, so their attaching spheres can be drawn in the same 3-D diagram.} to further refine the exoticity and plausibly birth different species of elementary fermions, or link them up to engender bosonic interactions. Therefore, strictly speaking, no actual clean punctured (happened during the connected summing) $\mathbb{CP}_{\text o}^2\equiv {\mathbb{CP}}^2-\text{D}^4$ need to be present in our universe, and ``lairs'' must be used in a more inclusive sense to represent $\pm1$ framed 2-handles that may not be isolated. We won't be fastidious in the ensuing discussions though, since no confusion would likely arise.

On the other hand, we propose that the bosons are sprinkled (by the $T^2$-surgery) mostly over the remaining $S^1\times S^3$ components, but can clasp onto the fermionic liars and bridge them (cf., the boson lines in Feynman diagrams). Since bosons don't repel each other and can happily overlap, we should not need a separate $S^1\times S^3$ to host each boson, but instead, all the bosons could cohabitate in the same overall macroscopic $S^1\times S^3$ universe\footnote{Strictly speaking, the decomposition of a spacetime into a minimal (not the blow-up of any other manifold) base and collection of lairs is not unique, meaning that the minimal manifold obtained from a sequence of blowing-down operations (shrinking $-1$ framed 2-handles to single points, dragging along other handles that were initially linked with it to become linked with each other; strictly speaking, only the particles get blown-down, we abuse terminology and let blow-down here to also refer to the removal of the anti-particle lairs) depends on the order in which the lairs are eliminated. However, geometry, specifically the existence of metrics and thus length scales, that allows for a distinction between the macroscopic and the microscopic, offers a preferred choice for a macroscopic background base universe, which is what we are discussing here.}. Besides, the requirement for multiple copies of $S^1\times S^3$s in the generic construction of arbitrary manifolds is due to the need for manufacturing nontrivial fundamental groups, where each copy corresponds to a generator of the group, thus having more of them just leads to more complicated multi-connectedness, that could physically manifest as a plethora of Aharonov–Bohm type of effects, which our universe does not appear to observationally exhibit. In any case, in our proposal, we aim to model particle physics as nontrivialities in smoothness, not homotopy, so a multitude of $S^1\times S^3$s is not really economical in terms of satisfying the needs of theoretical model-building either\footnote{In fact, the study on exoticity tends to concentrate on simply-connected manifolds. So the inclusion of many $S^1\times S^3$ factors will compress the available pool of results that we can draw on. Our proposed macroscopic universe, even with just one factor, is already slightly more complicated than usual, since $\pi_1(S^1\times S^3)=\mathbb{Z}$, so some extra technicalities (such as 2-torsion in the second cohomology group) need to be considered. In any case, since we cannot or don't expect the physical theories to sail through the big bang (a $S^3$ slice) without glitch, it is acceptable if quantities become ill-defined at places, so we really only need to deal with a contractable region (the observable universe), with the benefit of e.g., having a uniquely defined spin structure.}. 

Moreover, an toroidal $S^1\times S^3$ universe matches the metric-signature-switch explication of the big bang in \cite{galaxies8040073,Zhang:2019rrc}, where the switching transpires at the poloidal seam separating the inner and outer halves of the torus. Note though, the toroidal universe also admits a big crunch, which could be realized if dark matter is indeed explained by autonomous Weyl curvature as proposed by \cite{galaxies7010027}, in which case the growth in entropy (contributed mostly by the vast gravitational degrees of freedom \cite{1989NYASA.571..249P}) over time will compel a persistent creation of additional dark matter that eventually overpowers dark energy. Finally, we note that for an intersection form\footnote{Closely related to the linking matrix of 2-handles (cf., Proposition 4.5.11 of \cite{KirbyCal}).} $\mathcal{Q}$ with only $\pm 1$s on the diagonal, which is appropriate (see Exercise 1.2.10 through Theorem 1.2.21 of \cite{KirbyCal}) for a closed ($\mathcal{Q}$ is unimodular) braneworld universe like the one we are adopting here, containing at least some antiparticles ($\mathcal{Q}$ is indefinite), the difference between the particle and antiparticle populations of fermions is also the signature of $\mathcal{Q}$, which is often strictly regulated\footnote{It equals the difference between the second betty numbers $b_2^+-b_2^-$ of the braneworld spacetime, and is related to the Pontryagin class of its tangent bundle, as well as the Chern numbers when a complex structure is available.}, e.g., it must be divisible by $16$ for a smooth, simply connected, closed, oriented manifold. More interestingly, it has to be vanishing for a closed, oriented smooth manifold that embeds into $\mathbb{R}^6$ (which is more relaxed than our embedding into $\mathbb{R}^5$, thus is a necessary condition). This would imply that fermionic particles and antiparticles must equalize in number as far as the \emph{entirety} of the universe is concerned, which suppresses CP violation, thus provides a possible avenue for solving the strong CP problem. Note though, this vanishing signature is a global topological constraint, so some local fluctuations for our nearby observable universe is not prohibited. Indeed, the very small observed baryon asymmetry parameter values, $9$ to $10$ orders of magnitudes smaller than its natural value \cite{2018arXiv180110059S}, may arguably be of a perturbative origin.

\subsubsection{Differential identity} 
The exoticity-generating surgeries are concocted by cutting out a submanifold, and re-glue it back after applying an involution on its boundary\footnote{Or glue back something different but share the same boundary thus also slots into the open socket.}. When the boundary involution\footnote{While higher order periodic diffeomorphisms are possible \cite{2019arXiv190202840M}, involutions are sufficient for exoticity, e.g., \cite{1997math.....12231K} only utilizes involutions.}, after being extended into the interior of the removed piece, preserves its topology but not the smoothness, it is called a cork\footnote{That is, it is only relatively exotic -- it doesn't lack a diffeomorphism into the standard smoothness structure, it just lacks one that restricts exactly into the prescribed involution on the boundary. The involution (included in the definition of the cork) determines how it glues onto the smoothness structure of the rest of the manifold, and ``awkwardness'' at the seam forces the piece to ``twist its posture into a pretzel'' to comply. This contortion is obviously more severe in differential structures than topological structures since the requirement of smoothness prevents the unwinding of the stressing via buckling, which tends to be accompanied by the appearance of singularities like fold lines. Consequently, it is not surprising that the extension of the involution can be a homeomorphism while failing to be a diffeomorphism.}, and if both are changed, it is called a plug \cite{2008arXiv0806.3010A}. These special cases have garnered much attention, because they have been witnessed in various example cases, and any simply-connected (situation with other connected cases is unproven yet) closed 4-manifold can be depicted using corks (in fact, a single, perhaps complicated cork) alone \cite{Mat,CFHS,AK1}. In other words, while one can cook up all sorts of innovative surgeries to create exotic smoothness, the same destination can always be arrived at by taking the corks route. In fact, it is further conjectured that a limited collection of elementary corks and plugs is sufficient for describing all exoticity \cite{2008arXiv0806.3010A}, which would be temptingly similar to the situation we see in the hierarchy of particles.  

We would then like to identify elementary particles with elementary corks and plugs, and it is with this additional constrain of only including the elementary types in our consideration, that we tighten the definition of corks and plugs in our discussion. On the other hand, we may have to also loosen their definition somewhat, since a naked lair, being a punctured $\pm\mathbb{C}\mathbb{P}^2$, is a tubular neighbourhood of a sphere of self-intersection $-1$ (see proposition 2.2.11 in \cite{KirbyCal}), which is not allowed to be present in a Stein manifold. There is then possibly (although it is unclear what the additional handles sitting on top would do in this regard) an infringement of the Stein clause in the definition of corks and plugs in \cite{2008arXiv0807.4248A}. But this Stein pre-requisite is more of a crutch, while able to facilitate simpler proofs of exoticness\footnote{It collects some conclusions of the Donaldson \cite{10.4310/jdg/1214437665} or Seiberg-Witten \cite{SEIBERG199419} type of gauge theory investigations into more tractable packaging. Often, one can prove that two manifolds are non-diffeomorphic by showing that one is Stein and the other isn't.}, isn't really a necessary condition (e.g., it does not distinguish between smoothness structures differing only in Bauer-Furuta invariants) for generating exoticity (indeed, whether Stein-ness is included in the definition of corks and plugs differ by author), and thus shouldn't be expected to be present automatically in a physical context, so can be dropped to yield the so-called \emph{loose} corks \cite{2016arXiv160509348A, Mat} and plugs. From here on in this paper, we will be referring to the more general loose versions of these entities implicitly. 

But then which of corks and plugs jibes with fermions, which with bosons? Plugs tend to appear whenever corks get destroyed or undergo mutations, so their relationship is indeed reminiscent of that between fermions and bosons, although in the particle case, one can see either as the catalyst for changes occurring in the other, so this observation does not quite fix the correspondences. One then needs to dig deeper into the differences between plugs and corks. In this respect, we recall that besides whether involutions extends inwards continuously in their definitions, the corks are contractible (i.e., homotopic to single points) while the known plugs are homotopic to $S^2$. Such nontriviality in algebraic topology seen with the plugs\footnote{It is also one of the motivating factors prompting us to encase the fermions in lairs. As the fundamental building blocks for nontrivial topology in four dimensions (see footnote \ref{fn:legos}), it would be extremely surprising (i.e., fine tuned) if there aren't any of these lairs in our actual physical universe. And if they are present, they would more likely to be concomitant with plugs.} must be confined inside small spatial regions and hidden from sight beyond the energy scale we can probe the particles with, or else we would have observed all sorts of (including but not restricted to gravitational) consequences of a nontrivial spacetime topology. In other words, the bosons, which can be rather expansive (take for example low frequency radio waves), are not very likely to be the plugs, so we identify corks with bosons\footnote{This implies that if there can exist a universe without any fermions but contains bosons, which don't tend to complicate algebraic topology, we might end up obtaining an example of an exotic $S^1\times S^3$ (there is a statement in \cite{Brans:1994hq} that such a manifold likely exists), which would be mathematically interesting, since simpler exotic manifolds like $S^1\times S^3$, $S^2\times S^2$ and $\underline{\mathbb{CP}}^2$ are stepping stones towards proving the smooth Poincar\'e conjecture.} and plugs with fermions\footnote{Note, our assignment is the opposite to an analogy evoked in \cite{2008arXiv0807.4248A}. Also, the ``positron'' move in the differential topology literature, that makes the Thurston-Bennequin number more positive and thus the manifold closer to Stein, probably has no direct counterpart in physical particles. Furthermore, an anticork (a piece of a cork obtained by carving out a disk), which being relatively fake and homotopic to $S^1$, is very different to a cork, so has nothing to do with antiparticles. The mathematical and physical nomenclatures diverge in general.}. 

\subsubsection{Spin statistics}
To further justify this assignment, we note that with plugs, the interior extension of the boundary involution $\tau$ is extricated from the straightjacket of continuous invertibility, thus is possibly endowed with the necessary flexibility to accommodate the non-integer spins of the fermions. To see how, let $W$ be a cork or plug that attaches onto some outside manifold $K$, then prior to gluing the two pieces together, we can lay a vanilla coordinate system (atlas) $\tilde{x}_W$ inside $W$ and a coordinate system $x_K$ inside $K$. The two sides then cannot be extended (into collars beyond the boundary $\partial$) and glued smoothly together (i.e., they don't admit a smooth transition function in the overlapping collars) once $\tau$ is taken into account and the combo becomes exotic. There is however a different, exotic, atlas $x_{W} = \tau(\tilde{x}_{W})$ (recall $\tau$ is a homeomorphism for corks and not so for plugs) inside $W$ that can be glued onto $x_K$ smoothly. Unfortunately, because our present particle theory fails to appreciate exoticity, thus assumes the simplest but wrong vanilla $\tilde{x}_W$ (and subsequently the holonomic spinor basis) when writing down expressions, we will have to figure out how coordinate transformations on $\tilde{x}_W$ behave as well. In particular, when we carry out a $\varphi$ [e.g., $\in \text{SO}^{+}(1,3)$] transformation to the outside coordinates $x_K$, it extends smoothly onto $x_W$ (for brevity, we will leave implicit the smooth transition functions, either across $\partial$, or between charts on the same side), but won't do so for $\tilde{x}_{W}$. What it will see instead, is a transformation $\tilde{\varphi}$ satisfying $\tau \circ \tilde{\varphi}(\tilde{x}_W)=\varphi(x_{W}) = \varphi\circ \tau(\tilde{x}_W)$, thus using $\tau \circ \tau |_{\partial}= \text{id}$, we obtain that $\tilde{\varphi}|_{\partial} = \tau \circ {\varphi} \circ \tau^{-1}|_{\partial}$. 

This relationship is not particularly interesting if $\tau$ extends into the interior of $W$ as a one to one homeomorphism as in the case of corks, but with plugs, it could become many-to-one and non-invertible\footnote{One can appreciate that the involution is a more drastic action for the plugs than for the corks, by noting that a typical $\tau$ would be a dot (1-handle)-zero framing (2-handle) switch, or reflection against a symmetry axis on the Kirby diagram. In the cork case, there typically exist many invariant loops that can sneak back to their pre-involution initial position by isotopy, i.e., continuous shifting of the lines in the diagram. With plugs on the other hand, too much stuff is in the way for many such loops to exist, so their Kirby diagrams have been rearranged more radically.}. Then if we formally (recall that we are oblivious to exoticity, thus also the fact that it is not legitimate to do so) carry this relationship off of $\partial$ and into $W$, the $\tau^{-1}$ would become ambiguous and could provide two or more different targets for $\varphi$ to act on, yielding several different results (the further action by $\tau$ won't necessarily reverse this effect, since the multiple outputs of $\varphi$ do not need to be among the pre-images of the same output for $\tau$ anymore -- $\varphi$ and $\tau$ are completely independent transformations, thus $\varphi$ is under no obligation to preserve the degeneracy classes of $\tau$). In other words, there could be multiple $\tilde{\varphi}$s corresponding to the same $\varphi$, and when the multiplicity is two, we could have $\tilde{\varphi} \in \text{SL}(2,\mathbb{C})$ double-covering $\varphi \in \text{SO}^{+}(1,3)$. 

Heuristically, the issue underpinning spin statistics is thus the difficulty in extending $\tau$ continuously inwards. In particular, the source of the indecisiveness of the projective representation of the Lorentz transformations by spinors is rooted in the ambiguity of $\tau^{-1}$. 
Recall that $\tau$ is a diffeomorphism on the boundary but cannot be propagated smoothly into the interior of $W$. If we try, problems necessarily develop when various fronts of continuation off of different sectors of the boundary end up meeting each other. In the case of plugs, we encounter more problematic singularities, e.g., the coordinate lines of $\tilde{x}_W$, when seen from the perspective of the more appropriate $x_W$, can develop caustics, so that the Jacobian (derivatives of $\tau$) between the two coordinate systems become non-invertible, and such caustics can further serve as the branch points of a multi-valued $\tau$ inside of $W$. 

\subsection{Fields} \label{sec:fields}
The details of the corks and plugs are clearly not being recited by the presently available particle models, so a hefty dose of coarse-graining is involved. However, we are not arriving at the coarse Standard Model description in a top-down fashion, by integrating over more detailed dynamics as in statistical mechanics, but instead from below, by trying to reverse-engineer it via a bottom-up compromise, starting from even coarser phenomenological bookkeeping expressions. It should not then be surprising, that the infrastructure being cobbled together would look somewhat labyrinthine, once re-interpreted from the top-down perspective that our exotica proposal aspires to bring forth. 
A central piece of this re-interpretation endeavour is the strategy for subsuming the lairs into the field theory language, or in other words, the manner in which lairs empathize with the spinor bundle. A sketch of one possibility of how this might work is as follows:
\begin{itemize}[labelwidth=!,labelindent=0pt]
\item The spinor bundle fibre (a complex 2-component vector space) over each macroscopic $S^1\times S^3$ spacetime location $q$ is in fact the tangent vector space of a base point $\hat{q}$ in a liar, being transplanted over to become exposed to the outside world (note since the lairs are $\pm \mathbb{CP}_{\text o}^2$, we automatically have $T_{\hat{q}}(\pm\mathbb{CP}_{\text o}^2) \cong \mathbb{C}^2$ without effort). We need now to locate this original base $\hat{q}$.

\item When there is indeed a fermion, the worldline (spatially pinched throat between the $S^1\times S^3$ and the lair) of which passes through our $q$ of interest, the point $\hat{q}$ would then rest on a true ambient $\mathbb{R}^5$ geodesic\footnote{As opposed to just a geodesic of the embedded spacetime, which is the shortest path if we are confined to the spacetime, while ambient geodesics are the shortest even if we are allowed to lift off of the spacetime, and exploit the additional freedom thus afforded.} threading through the interior of the lair. In other words, the tangent planes along an ambient geodesic inside of the lair, and in particular the moving frame bases (we will later argue that the Dirac spinors constitute fragments of a Darboux-esque \cite{Darboux1915} moving frame) contained within, get transplanted onto the exposed worldline, which then serves as a surrogate or representative of that hidden ambient geodesic. 

Such ambient geodesics are particularly informative about the lair geometry (see Sec.~\ref{sec:equations} below), just as the null rays of the macroscopic $S^1\times S^3$ (also ambient geodesics \cite{galaxies8040073}) dictate the vital causal structure there. Therefore, when our modeling efforts are eviscerated by our ignorance into a coarse-grained description, where information about the four dimensional lair region's geometry has to be condensed into being portable by mathematical concepts defined along a one dimensional worldline, the optimal strategy\footnote{An alternative is to integrate over spatial slices of the lairs, but integrations usually average over, thus smear out the geometric details, leaving us with only global topological information, yet we want geometry since we are interested in the energetics.} is to transcribe over the information along these similarly one dimensional sampling curves that best probe the said geometry. 

\item 
When there isn't a lair passing through $q$, the spinor fibre over $q$ in the spinor bundle isn't transplanted from anywhere, it is just a dummy fibre wherein the moving frame vanishes so the fields take zero value. It is merely a padding that facilitates formally extending quantities defined on a worldline into the entire spacetime, and should be used in conjunction with Dirac delta distributions, multiplied into, e.g., the amplitudes of the Dirac spinors. 

In practice though, a further randomization procedure will be applied (see Sec.~\ref{sec:equations} below), so the Dirac delta distributions (in the sense of generalized functions) are often replaced by smoother (in the vanilla differential structure) probability distributions, and thus in some cases\footnote{\label{ft:dist}For example, the classical solutions of quantum fields. This smoothing is not always possible or perfect though (e.g., right after a measurement, some smearing by ignorance gets taken out and we are hit once again with a delta distribution in the measured observable), and quantum theories generally cannot do away with generalized functions. One should perhaps suffer this state of affairs with some angst, since multiplying generalized functions is mathematically rather tricky (cf., Colombeau algebra \cite{Colombeau}), thus the depiction of quantum interactions is inevitably suspect.} the fields become differentiable not just in the weak sense, allowing calculus manipulations to proceed more fluently. In other words, we superpose, in the fashion of Feynman path integral, all possible lair configurations (assuming single lair in the case of quantum mechanics, and a mutable number of lairs in the case of quantum field theory), so the spinor fibre over $q$ becomes the probability-weighted average of tangent planes at many $\hat{q}$ of all different possible configurations. 

\end{itemize}

\subsection{Mass} \label{sec:mass}
With fermions, the impact of mass manifests as a phase rotation of the spinors (most plainly demonstrated by particles at rest, in the Dirac representation), which would appear rather natural if the two halves of the Dirac spinor represent two complex moving frame bases that are being transported along a sampling ambient geodesic that spirals into a helix inside of the lair\footnote{In fact, $\mathbb{CP}^2$, being nearly the direct product between a torus and the positive octant of a sphere, can have one of its toroidal ``fibers'' being ruled by such ambient geodesics, that wrap around the torus (the wrapping trivializes for those singular ``fibers'' that are squeezed into circles) into helical shapes.} \footnote{An $\mathbb{R}^5$ ambient physicist on the other hand, would report a dual perspective, where it is the sampling ambient geodesic that appears straight, while the lair region of our braneworld spacetime spirals around it.\label{fn:spiral}}. As the toy model for a chiral fermion, take for example a circular helix, for which the important parameters are the radius $r$ and pitch $\hat{p}$ (we also define $p\equiv \hat{p}/2\pi$ for convenience), the slope $s=p/r$, as well the frequency of phase rotation $1/\hat{p}$, which can be compared with that of the Dirac spinors, $2 \gamma mc^2/h$. The entries in the latter expression, apart from mass, are all generic and not specific to particular particles: 
\begin{itemize}
\item 
$\gamma$ being the time dilation is simply a calibration factor for the longitudinal temporal coordinate used to measure the pitch of the helix against, it is included so the time that multiplies onto this frequency gets adjusted to the comoving time. 

\item 
$c$ is, as usual, just a scaling ratio that synchronizes the temporal and spatial units of measurement. A relic of the pre-relativity era, we set it to unity from here on.

\item 
$h$ is a normalization factor synchronizing the mass and temporal units (with $c$ slotted in where appropriate), or in other words, by dividing into a time interval, it translates the temporal units, like seconds, that's more familiar from our daily lives, into multiples of units more typical for the temporal (and spatial, via $c$) scales associated with elementary particles. This is why it appears in places like the uncertainty principle, demarcating the temporal-spatial length scale at which we enter the microscopic world and has to enrol quantum theories. 

Specifically, let us first assume $\gamma \approx 1$, so non-relativistic quantum mechanics and thus the elementary form of Heisenberg's uncertainty principle applies. Then note that with the helix picture, the speed $v$ of the particle is simply the projection, onto the timelike longitudinal direction, of the null tangent vector to the helix (recall that this is a true ambient geodesic like the null rays), given by $v \approx c s =  p/r$ (where we used $c=1$ and $\hat{p}\ll r$ from the non-relativistic condition). It is furthermore reasonable to let $\Delta x \geq r$, since we cannot hope to pin down the point-like approximation of a particle into a region smaller than even the helix, because we would already be seeing into the gut of that particle's handlebody and the coarse-graining makes no sense anymore. It is also reasonable to assume $\Delta v \sim v$ so that the perturbations to the tightness of winding is of natural sizes. Then  our expression for $v$ becomes $\Delta x \Delta v \geq p$, and subsequently $\Delta x\Delta P \geq \hbar /2$ ($P \equiv m v$) once we substitute in $\hat{p} = h/(2m)$ (equating the helix frequency with spinor rotation frequency), reproducing the most commonly encountered uncertainty relation.

\end{itemize}
This leaves $m$, that does vary from particle to particle, to correspond to the inverse of the pitch, that prescribes the absolute tightness of the helix' winding. This identification has the following immediate implications:
\begin{itemize}

\item When $m=0$, the helix doesn't spiral and instead regress into a straight line inside the lair, implying that the lair's longitudinal direction, tangential to the particle worldline, is already aligned with an ambient geodesic. Because the worldline is exposed to the macroscopic world, it must then be a null ray. 

\item Fermions come in three generations, differing only in mass. Given that mass is geometric and not topological (either algebraic or differential) in nature, so should the generations, as possibly three solutions to the equiaffine isoperimetric equation, i.e., equiaffine mean curvature equaling the cosmological constant, which can be approximately set to zero for most problems relating to high energy experiments. In the case of massless neutrinos, we have a degenerate three dimensional solution space\footnote{\label{ft:ParticleSpinor}Note that, unlike $\text{SL}(2,\mathbb{C})$, gauge transformations like weak isospin $\text{SU}_{\rm iso}(2)$ do not change the spinors' directions (in the sense of being square-roots of null vectors), and instead act on an abstract space with a dyad basis formed by piecing together two Weyl spinors yanked out of two different particles. The 2-component complex vectors in this space (the Higgs doublet being one example) can at least be formally treated much like spinors (similar to the ones termed ``particle spinors'' in \cite{2017EPJP..132..446Z}). The distinguishing characteristic of such a dyad is that its bases solve equations of motions, and are thus geometrically meaningful entities that don't really change when the coordinates are altered, so vectors written under it, like the Higgs, form singlet representations of the Lorentz transformations. In a similar fashion, Weyl spinors yanked out of three neutrinos of different flavors can be grouped into a special ``particle triad'' for the degenerate solution space in the main text.} that can serve as the domain of the projection operator in \cite{2017EPJP..132..446Z}, while the lack of potential barriers separating solutions encourages flavor conversions even during free propagation.   

\end{itemize}

Without knowing much about those particle-specific handles residing in the lairs, these solutions of isoperimetry cannot be derived \emph{ab initio}, so the phenomenological mass values will have to be injected by hand, and the agent for smuggling them in is the Higgs field. The reason why the Higgs value is mutable is because the geometric details like mass would not just depend on a fermion's handlebody, but also the geometry of its immediate vicinity, in particular, whether the fermion plug is intersecting a boson cork, which exerts its influence by contributing to the covariant derivatives in the Higgs' kinetic term.
Moreover, beside bosons, there are also other factors competing to sway the Higgs values, e.g., the Yukawa coupling terms with fermions announce the impact of the presence of fermions on the local geometry. More significantly, there is also a resistance to (equivalently, energy cost associated with)  deviations away from the background pre-set vacuum expectation value, encoded in the Higgs potential, as well as the kinetic term for the Higgs. Respectively, the Higgs potential strongly penalizes geometric disturbances impacting the Higgs amplitude, while the kinetic term further raises the cost of any local (inhomogeneous) excitations even in the symmetry breaking directions of that potential\footnote{When boson corks are present to alter the local conditions however, it becomes energetically favorable to have some of those excitations, since it is the covariant, and not the partial derivatives that we want to suppress. The Goldstone modes are thus intimately affixed to the bosons, and it is appropriate, not only in mathematics but also in heuristics, that they be eaten and considered in conjunction with the bosons.}. 

Therefore, the coupling between the bosons and the Higgs through the covariant derivatives represents not an absolute authority of the bosons on the Higgs, but rather the rope in a tug-of-war against all these other factors, transmitting also their clout back onto the bosons.
A metaphor would be when a river (boson) carries a fallen log (Higgs), but the log becomes pinned by a rock (other factors) on one end, so the water ends up having to shift its flow streams and go around the log, but not without pushing the log into an angle causing least possible blockage, so the final state of affairs is a compromise between multiple elements. In other words, the quadratic coupling term from the covariant derivatives, that ultimately bestows the mass term on the bosons, signifies a scattering of the boson propagation by the partially stuck Higgs. Boson masses are thus of a different geometric origin to the fermion masses, fittingly introduced into the Standard Model in an ostensibly distinctive manner from the latter.  

\subsection{Charges}
The gauge symmetries are vital for the construction of the particle theories, in that it tells us how to lump fields together into interaction vertices (that must be gauge invariant) in a Feynman diagram. 
In our exotica context, these interaction vertices resolve (cf., point-splitting regularization; there are no divergences if exotica is accurately reckoned with, since everything is smooth) into hub regions where different elementary plugs and corks join, e.g., through linking up the attaching spheres of some of their handles, to eventually form a very complicated handlebody that is our universe, containing a vast number of interacting particles. The gauge invariance of the Lagrangian should then be corporealized by the constancy of the equiaffine volume element (the isoperimetry thus manifests as the extremization of the action), when certain handle manipulations are applied to those particles participating in the interaction. 

When we have at hand a detailed Kirby diagram, showing all the individual primary handles, such manipulations would be revealed as no-drama smoothness-structure-preserving\footnote{Rather apt for the prefix ``gauge'', although not rooted in coordinate freedom, but instead freedom in how one draws the Kirby diagram.} handle moves, that only alter geometry. One then immediately appreciate that the physical desirability of gauge invariance is simply to avoid fine tuning, to ensure that small perturbations to the postures of the participating particles will not instantly disqualify an interaction (due to non-compliance with the isoperimetric condition), thereby imbuing a level of robustness that allows repeated experimental verification. However, we do not know of the detailed handlebody constituting a particle, so we have to rely on abstraction to sweep our ignorance under the rug. Namely, we assign labels to particular (not explicitly known) configurations of hordes of handles, by e.g., giving them names like electrons, or other sub-designations like the color of a quark. Then as the aforementioned gauge manipulations revise the apparent relationships between the primary handles, their grouping monikers need to acclimate accordingly, and therefore the abstract label spaces inevitably harbor representations of the gauge groups. 

The charges of a particle (i.e., a composite handle configuration, including stipulations on geometry), which index those representations, then simply unveil into indicators of the composite handle's symmetries under the kindred manipulations. For analogy, an axisymmetric object would embody a singlet representation of rotations around the axis, thus appear uncharged. On the other hand, a $\pi$-symmetric object (i.e., the stabilizer subgroup being $\mathbb{Z}_2$) would have $\pi$ instead of $2\pi$ as the period of rotation, so it would pick up a phase increment of $2\Delta\phi$ when the rotation only advances $\Delta \phi$, or in other words, one ends up with a charge of $2$ (the cardinality of the stabilizer).  
Accordingly, within the context of exotica, the quantization of the charges could simply be a consequence of the cyclic discrete symmetry groups being indexed by natural numbers, instead of due to their being some sort of topological indices, which had been the more popular approach with previous geometrodynamic literature (see e.g., \cite{Atiyah:2011fn}). 

\subsection{Equations} \label{sec:equations}
Since the Dirac-Yang-Mills equations govern the details of the particle motion and energetics, they are geometric and not differential topological in nature. It is almost conspicuous\footnote{Simply compare the generic non-abelian gauge transformation for a fermion $\psi$ and boson $A_{\mu}$ 
\bea
\psi \rightarrow  U\psi\,, \quad A_{\mu} \rightarrow U A_{\mu} U^{-1} + \frac{i}{g}(\partial_{\mu}U)U^{-1}\,, 
\eea
with a change of moving frame basis $X$, together with associated changes to the connection form $\omega$
\bea
X \rightarrow a X\,, \quad \omega \rightarrow a^{-1} \omega a + a^{-1} d a \,.
\eea
Note the differing common conventions for the suppressed matrix multiplication indices in the two disciplines is the reason for the slight difference in the ordering of the terms.} that their progenitors are likely to be the structure equations, governing the evolution of moving frames. However, these equations are 
\begin{itemize}
\item \emph{Redacted}: as discussed in Sec.~\ref{sec:fields} above, the moving frame associated with e.g., a solitary Dirac fermion, is attached to a single sampling curve but transplanted onto the worldline. Such a moving frame would be similar in functionality to the Darboux frame, but with an asymptote (ambient geodesic) replacing the principal curve (of the spacetime's embedding into $\mathbb{R}^5$), as the underlying carrier. In any case, the carrier curve is still specially adapted to the extrinsic geometry of the spacetime hypersurface, and thus, like the Darboux frame, its adapted moving frame has been sifted to more cleanly represent the geometric details of the spacetime, uncontaminated by the esoteric specifics of the sampling curve choice. In the more formal mathematical lingo, this means the spacetime geometry data being thus conveyed are well-defined, in the sense of possessing utmost invariance against alterations in the constructive crutches that appear in their definitions. 

Take a 2-D surface theory analogy, the geodesic and normal curvatures would vanish for an ambient geodesic's adapted moving frame, leaving only the torsion term (commensurable to mass in our context) to depict the twisting of the surface around the ambient geodesic (cf., footnote \ref{fn:spiral}). This jettison of irrelevant trivia also facilitates a more succinct formalism, whereby an abridged (e.g., two Weyl spinors pointed along the asymptotic directions; or even just one if they decouple and the other adds no useful information) version of some (properly 5-real-piece) moving frame, written from the intrinsic spacetime dweller's, rather than the usual ambient observer's dual, perspective, already suffices for the purpose of mapping out the salient geometric features of the lair regions of our spacetime.  

In summary, we do have a sensible choice of frame. However, a proper moving frame, and the structure equations governing it, should cover all the crevices of the handlebodies, in order to fulfill their duty as integrability conditions ensuring the embeddability of the exotically smooth spacetime into the $\mathbb{R}^5$ ambient. Ignoring the internal details of the lairs reduces the equations to necessary but not sufficient conditions. It also engenders a more practical problem, namely the derivatives in the structure equations should be taken against coordinates inside the lairs, in the vicinity of the sampling curve, yet only the post-pinching macroscopic spacetime coordinates are available to us, so a constructive approximate coordinate mapping had to be enlisted, in the form of gamma matrices or soldering forms, that enables us to transplant (besides the moving frame itself) also the derivative operators from those acting on the sampling curve into counterparts acting on its worldline surrogate. Because the two curves are actually quite different in their geometry, this formally grafted (now Dirac) equation, while still retaining the basic linear cross-coupling appearance (thanks to the linearity of the transcription procedure), becomes rather difficult to interpret -- the parameters appearing in it, like mass, don't equate directly to any curvature of the worldline, thus one would not immediately think of it as a structure equation. Only when we uncloak the original sampling curve, would the geometric import manifest (cf., the discussions in Sec.~\ref{sec:mass}). 

With the Yang-Mills equations on the other hand, while it is explicitly a structure equation governing the evolution of connections, these connections (such as weak isospin) don't act on the foregoing fundamental Darboux-esque moving frame. They act on the ``particle dyad'' instead (see footnote \ref{ft:ParticleSpinor}, and also the ``label space'' discussion in the previous section), which is an abstract moving frame constructed by blending Darboux-esque bases associated with two different geometries, and thus describes some convoluted hybrid (superimposed) geometry, which is a convenient bookkeeping tool to track the conversion between the geometries without needing to provide a detailed description of exactly how. In other words, the geometric (and in fact even the coarser handlebody) details within the lairs are heavily redacted, not only for freely propagating fermions, as discussed in the previous paragraph, but especially during interactions.

\item \emph{Randomized}: the worldlines are smooth curves under the actual exotic differential structure of the spacetime, and when mapped into the vanilla side, via a homeomorphism that by definition cannot be a diffeomorphism, such exotically smooth curves must appear continuous but non-differentiable, thus rather erratic when looked closely, and impossible to precisely predict\footnote{This predictability crisis may be further exacerbated by the redaction-induced incompleteness in initial data.} using differential equations of motion like Dirac-Yang-Mills, drafted assuming vanilla smoothness. 

The fortunate other side of the coin is that such wilder fractal-like paths are friendly to statistical mechanics. For example, the ergodicity theorem requires the phase space trajectory of a particle to fill multi-dimensional regions, which is not possible if that trajectory is some smooth one dimensional curve, but becomes plausible if it has a higher (than the topological dimension) Hausdorff dimension, which can equal the embedding dimension (see Sec.~3.6 of \cite{Embedding}). Therefore, by borrowing some prowess from statistical mechanics (most explicitly via the Feynman path integrals, and perhaps also via relating the Heisenberg picture observables in canonical quantization to distributions in the sense of generalized functions), quantum mechanics (Dirac-Yang-Mills equations in particular) circumvents the differentiability deficiency by not evolving single worldlines, but instead their probability smeared smoother distributions. However, the wild particle trajectories are not simple martingale random walks like Wiener processes, since the fakeness on na\"ive spatial slices dissipates when we are far away from the particle (fermionic handles forcing exoticity are lair-bound), indicating that the source of stochasticity and the particle motion are closely intertwined. Trying to mimic such behavior presents challenges to statistical mechanical treatments, e.g., it is an on-going battle to render the integration measure for the Feynman path integral mathematically rigorous. 

It had long been noted that the Standard Model of particle physics is written in the differential geometric language of fibre bundles, yet it isn't clear what is the ontological entity the geometry of which it describes. Within the exotica proposal (and geometrodynamic constructs in general), this entity reduces once again to the spacetime itself, which allows for interpreting the Einstein's equation as being of a purely geometric origin (e.g., as the equiaffine Gauss equation as proposed by \cite{galaxies8040073}). However, the matter stress-energy is randomized, but the gravitational Einstein's tensor isn't, so one has to synchronize them before establishing equality. The traditional approach is to try and randomize (i.e., quantize) gravity in an analogous fashion to particle theory, but that route has been technically challenging, and not guaranteed to eventually succeed given the desultory nature of the procedure (cf., e.g., footnote \ref{ft:dist}), which might give rise to self-consistency issues (e.g., general covariance prohibits identifying points in different warped spacetimes, thus prevents the linear superposition of gravitational quantum states \cite{1996GReGr..28..581P}). A more viable and satisfying alternative might be to try and de-randomize the particle side instead, by acknowledging exoticity, and summoning the actual detailed structure equations written under the correct differential structure. 
\end{itemize}

\section{Conclusion}
In this brief note, we speculated on some aspects of particle behavior that may be explainable by our spacetime possessing exotic differential structures. In fact, we are not aware of any argument why Nature would prefer the vanilla option, which constitutes a null set out of all available choices, so the question is perhaps more ``why not'' rather than ``why''.  After all, mathematics is the language of natural philosophy, enabling the rigorous logical deductions that power our search for a comprehension of the physical world. So when it offers a vast new lexicon, it would be insanity not to take advantage, especially since we are currently sputtering so badly when trying to narrate the microscopic world. In the reverse direction, should the proposed connection be confirmed, the accrued physical knowledge from particle experimentation offers a deep reservoir of reliable facts and clues to a differential topologist. 

Despite the obvious appeal of exotica though, we caution that the viability of such a proposal is difficult to quantitatively assess at the moment, due to the technical hurdles one encounters when dealing with low dimensional differential topology. When even the intensely scrutinized four dimensional smooth Poincar\'e conjecture is still outstanding, we cannot possibly hope, for the moment, to clarify (in increasing levels of detail) the topology, smoothness structure, complex structure, as well as the Euclidean and equiaffine geometries of a generic four dimensional spacetime manifold, that would be necessary for such a re-interpretation of the Standard Model of particle theory. What we can hope for here, is at its most ambitious, a synopsis of which caverns might have potential for spelunking fun. In particular, the more constructive aspects of the discussion are only meant to demonstrate possibilities, and we expect major overhauls once in-depth investigations become tractable.  

\newpage
\acknowledgements
This work is supported by the National Natural Science Foundation of China grants 12073005, 12021003, 11503003 and 11633001, the Interdiscipline Research Funds of Beijing Normal University, and the Strategic Priority Research Program of the Chinese Academy of Sciences Grant No. XDB23000000.

\bibliography{Uncertainty.bbl}

\end{document}